\documentclass{elsarticle}

\usepackage{lineno,hyperref,amsmath,amssymb}
\usepackage[numbers]{natbib}
\begin{document}

\begin{frontmatter}

\title{Effective medium approach in the renormalization group theory of
phase transitions}

\author
{V I Tokar}

\address{Universit\'e de Strasbourg, CNRS, IPCMS, UMR 7504,
F-67000 Strasbourg, France}
\ead[url]{tokar@ipcms.unistra.fr}

\begin{abstract}
	An effective medium approach similar to the coherent potential
	approximation (CPA) in the theory of disordered alloys and to the
	DMFT has been extended to the renormalization group equations
	in the local potential approximation (LPA).  Non-universal
	characteristics of the second order phase transitions such as the
	critical temperatures and critical amplitudes have been calculated
	in good agreement with the best known estimates.  A possibility of
	cluster extension of the LPA to improve its accuracy and to make
	the approach systematic and self-contained has been discussed.
	A qualitative explanation of the discrepancy between theoretical
	value of the critical exponent $\beta$ and recent experimental
	data on ordering in beta brass by the influence of non-universal
	contributions has been suggested.
\end{abstract}

\end{frontmatter}

%\linenumbers

\section{Introduction}
Phase transitions in many-body systems can be formally defined as the
points where the free energy is singular with respect to the parameters
entering the partition function, such as the external field and the
temperature.  Because microscopic Hamiltonians depend analytically on
the parameters, the calculations of the free energy by a finite order
series expansion in powers of the Hamiltonian or of its part cannot be
singular. This means that phase transitions can be described only within
non-perturbative approaches capable of calculating the free energy to
all orders in the Hamiltonian which in general is a difficult task.

The problem simplifies in lattice models where the infinite system can
be modeled by a finite cluster of lattice sites.  In classical models
which will be discussed in the present Letter the cluster partition
function can be calculated exactly to all orders in Hamiltonian and
with a suitable embedding of the cluster into the infinite system
remarkably accurate results can be obtained with the use of small clusters
\cite{ducastelle,tokar_new_1997,maier_quantum_2005,tan_topologically_2011}.
The cluster approximation is systematic in the sense that it
can be indefinitely improved by enlarging the cluster size,
so in principle it presents a viable alternative to the series
expansions as a general approach to many-body problems with strong
interactions.  The main drawback of the cluster approach is that
computationally accessible dimensions of the clusters are limited
so it is impossible to adequately treat very long-range correlations
\cite{tokar_new_1997,tan_topologically_2011} which are indispensable
in the description of the second order phase transitions and critical
phenomena \cite{wilson}.

The critical behavior can be effectively described
within the renormalization group (RG) theory
\cite{wilson,berges_non-perturbative_2002} which proved to be
efficient in the vicinity of critical points where interactions
between large-scale fluctuations weaken and sophisticated perturbative
techniques make possible precision calculations of the universal
quantities, such as the critical exponents and the critical
amplitudes ratios \cite{le_guillou_zinn-justin,RG2002review}.
Experimentally, however, mainly non-universal quantities are
measured, such as the critical temperatures and individual critical
amplitudes, but their theoretical calculation in strongly coupled
cases meets with difficulties.  Current non-perturbative RG approaches
\cite{BAGNULS200191,berges_non-perturbative_2002,latticeRG2010,RG2002review}
suffer from poorly controlled approximations and so are not reliable
enough to quantitatively predict non-universal quantities.  Presumably
because of this, phenomenological theories and the high-temperature
expansions of simple model systems developed before the advent of the
RG theory are still being invoked to interpret experimental data on the
second-order phase transitions (see \cite{beta-brass2016} and references
therein).

Taking into account that some non-universal quantities, such as the
critical temperatures, can be calculated with good accuracy within the
cluster method \cite{tokar_new_1997,tan_topologically_2011}, in the spirit
of the RG approach it would be natural to use some cluster technique to
account for the short-range fluctuations and resort to RG only for the
treatment of the long-range ones.  A step in realization of this approach
was made in \cite{latticeRG2010} where the authors have succeeded in
calculating a magnetization curve and several critical temperatures in
good agreement with the best known values.  This, however, was achieved
with the introduction of arbitrary parameters into the calculations which
casts doubts on the predictive abilities of the approach.  Besides, no
recipe was given of incorporation into the RG procedure of the notion of
the effective medium which is central in the existing cluster approaches
\cite{tokar_new_1997,maier_quantum_2005,tan_topologically_2011} was not
introduced in the in the RG scheme in \cite{latticeRG2010}.

However, sound though not rigorous arguments exist that it is the
self-consistency equations defining the effective medium underlie
the success of such renown methods as the coherent potential
approximation (CPA) and the dynamical mean-field theory (DMFT) in
the strong coupling regime (see the bibliography to the subject in
review articles \cite{elliott_theory_1974,maier_quantum_2005} and also
\cite{gamma_exp,masanskii_method_1988}).  Especially relevant to us is
the CPA because by means of the replica trick the disordered alloy model
can be reduced to a particular case ($n=0$) of the $n$-vector models
which are the subject of the present study.

The main aim of this Letter is to derive a renormalization scheme based
on the local potential approximation (LPA) \cite{local_potential,1984}
and a self-consistency condition for the effective medium
\cite{gamma_exp} formally the same as was used in the cluster approach
in \cite{tokar_new_1997,tan_topologically_2011}, though currently based
only on the single-site cluster so the scheme will be analogous to the
single-site approximations (SSA) such as CPA and DMFT.  It will be shown
that the implementation of the self-consistent LPA scheme (SC-LPA) makes
possible accurate calculation of non-universal quantities without use of
arbitrary parameters.  As a first physical application of the method,
semi-quantitative arguments will be presented that the non-universal
contributions may be responsible for the discrepancy between the universal
value of the critical exponent $\beta$ calculated within RG theory and
the values obtained in recent experiments on ordering in beta-brass
\cite{beta-brass2016}.
\section{Definitions and notation}
In this Letter we will deal with classical statistics of the $n$-vector
models which describe real continuous $n$-component lattice fields on
periodic $d$-dimensional lattices which interactions are described by
Hamiltonians of the following general form
\begin{equation}
H[{\bf s}]=\frac{1}{2}{s}_{\sigma i}\epsilon_{ij}{s}_{\sigma j}
+\sum_iH_I({\bf s}_i)= 
\frac{1}{2}{s}_\sigma\hat{\epsilon}{s}_\sigma+H_I[{\bf s}].
	\label{H}
\end{equation}
Here the first terms on the right hand side (r.h.s.) describes the pair
interaction between the fields $s_\sigma=\{s_{\sigma i}\}$ at sites
$i$ and $j$ which may be different while the interaction part $H_I$ is
assumed to be local to the sites. Small Latin subscripts denote integer
$d$-dimensional site coordinates, the Greek subscripts $\sigma=1,\dots,n$
refer to the $n$-vector components. The boldface quantities will denote
the $n$-vectors and also the $d$-dimensional momenta ${\bf
k}_m$ in the lattice Fourier transform over spatial coordinates, but
the integer $d$-dimensional lattice site vectors $i,j$ for simplicity
will not be boldfaced.  To farther simplify notation, summation over
repeated discrete (but not continuous) subscripts will be implicitly
assumed throughout the paper. Besides, vector-matrix notation for the
lattice coordinates will be used as, e.g., in the matrix of the pair
couplings $\hat{\epsilon}=||\epsilon_{ij}||$ in (\ref{H}). In the general
anisotropic case the couplings may depend also on $\sigma$ but in the
present paper only fully $O(n)$-symmetric Hamiltonians will be considered.

It is important to note that separation of $H$ into the quadratic and
the interaction parts in (\ref{H}) is not unique because an arbitrary
quadratic term can be added to the first part and simultaneously
subtracted from $H_I$ with the total Hamiltonian remaining unchanged. This
trick will be essential in the effective medium theory in the next section
but in (\ref{H}) it will be used only to impose on ${\epsilon}_{ij}$
the requirement that its Fourier transform behaved at small ${\bf k}$ as
\begin{equation}
	\epsilon({\bf k})|_{{\bf k}\to0}\propto {\bf k}^2
	\label{k20}
\end{equation}
which is convenient in implementation of the RG technique \cite{wilson}.

The partition function of the $n$-vector models is defined by the multiple
integral over $N$ values $\{{\bf s}_i\}$ ($N$ the number of lattice sites)
of field ${\bf s}$ as
\begin{equation}
Z[{\bf h}]=\int \prod_id{\bf s}_i e^{-H[{\bf s}]+{\bf h}_i\cdot{\bf s}_i}
\label{Z}
\end{equation}
where the source field ${\bf h}$ has been introduced so that the partition
function can be used as the generating functional of the correlation
functions (CFs) of the field.

By the linked cluster theorem the generating functional of connected
CFs is given by $\ln Z[{\bf h}]$.  Two such CFs will be needed in the
present study: the magnetization
\begin{equation}
	\label{s_av}
	{m}_{\sigma i} \equiv \langle s_{\sigma i}\rangle
	= \left.\frac{\partial\ln 
		Z[{\bf h}]}{\partial h_{\sigma i}}\right|_{\bf h=0},
\end{equation}
and the connected part of the pair CF $G^R_{ij}$ 
\begin{equation}
	\label{ss_av}
	G_{ij}^R= \left.\frac{\partial^2\ln Z[{\bf h}]}{\partial h_{\sigma i}
\partial h_{\sigma j}}\right|_{\bf h=0}
=\langle {s}_{\sigma i}{s}_{\sigma j}\rangle -
{m}_{\sigma i}{m}_{\sigma j}.
\end{equation}
Though in general case non-trivial pair correlations between all
field components may exist, the CFs diagonal in $\sigma$ defined in
(\ref{ss_av}) are sufficient in the fully $O(n)$ symmetric case that will
be considered in the present paper in the $n>1$ models.  All diagonal
CFs are equal due to the symmetry so the subscript $\sigma$ may be
dropped. The spontaneous symmetry breaking will be discussed only for the
Ising model with $n=1$ so the subscript would be also superfluous. In
(\ref{ss_av}) and throughout the paper superscript $R$ denotes fully
renormalized quantities.

The Fourier transformed CF (\ref{ss_av}) may be expressed through the
exact mass (or self-energy) operator $r^R$ as
\begin{equation}
	G^R({\bf k}) = \frac{1}{\epsilon({\bf k})+r^R({\bf k})}.
\label{GR(k)}
\end{equation}
This representation is convenient in approximate calculations because
unlike $G^R$ which in the vicinity of the critical point $r^R({\bf
k}\to0)\to0$ is strongly momentum-dependent (see (\ref{k20})), the mass
operator in many cases can be approximated by a momentum-independent
constant as, for example, in the CPA, in DMFT, and in the critical
region \cite{wilson}.
\section{\label{func-diff}Functional-differential formalism and the
self-consistency condition}
A general self-consistency equation for the mass operator
within the functional formalism is discussed in detail in
\cite{gamma_exp,tokar_new_1997,tan_topologically_2011,arXiv19} so below
only short explanations will be given to the formulas that will be needed
in subsequent calculations.

In strongly coupled models the calculation of the exact $r^R({\bf
k})$ is usually beyond the reach so one has to resort to approximate
treatments. Useful approximations adopted, e.g., in CPA and DMFT, can be
obtained by the neglect of the momentum dependence of the self-energy. In
the case under consideration this amounts to assuming that the exact $G^R$
(\ref{GR(k)}) can be approximated by the following trial CF
\begin{equation}
G({\bf k}) = \frac{1}{\epsilon({\bf k})+r}\approx G^R({\bf k}).
\label{G(k)}
\end{equation}
The last equality means that the momentum-dependent function $r^R({\bf
k})$ is approximated by ${\bf k}$-independent constant $r$. Obviously,
there is much ambiguity in doing this which can be used to adjust the
approximation to satisfy some condition. In the SSA, like CPA and DMFT,
$r$ is set to be the average $r\approx N^{-1}\sum_{\bf k} r^R({\bf k})$
over the Brillouin zone (BZ)which is equivalent to assuming that the
mass operator is diagonal in the lattice coordinates. From the ${\bf
k}$-space standpoint this means that all Fourier momenta are equally
important in the problem under consideration.

But in the critical region the most important are not local contributions
but those with the largest spatial extent, i.e., with the smallest ${\bf
k}$, so it seems natural to account for this by setting
\begin{equation}
	r\approx r^R({\bf k}=0).
	\label{r=rR}
\end{equation}
The adequacy of this approximation will be verified in explicit
calculations below but it can be immediately noted that $G^R\approx G$
presupposes that the critical exponent $\eta=0$ because at the critical
point $r=0$ and from (\ref{k20}) it follows that as $|{\bf k}|\to0$ $G\sim
1/{\bf k}^2$.  This deficiency makes the approximation poorly suited for
2D systems where $\eta$ has an appreciable value so only 3D models will
be treated in the explicit calculations in section \ref{numerical}.

Explicit self-consistency condition for $r$ similar to the CPA and DMFT
equations is obtained as follows.  First, the field-dependent partition
function (\ref{Z}) which is the generating functional of CFs can be
expressed through the generating functional of the scattering matrix $S$
as \cite{hori_approach_1962,vasiliev1998}
\begin{equation}
	Z[{\bf h}]=
	e^{\frac{1}{2}{h_\sigma}\hat{G}{h_\sigma}} S[{\bf h}\hat{G}]
	\label{Z-S}
\end{equation}
or, equivalently, 
\begin{equation}
\ln Z[{\bf h}]=\frac{1}{2}{h_\sigma}\hat{G}{h_\sigma}-U^R[{\bf h}\hat{G}]
	\label{lnZ-S}
\end{equation}
where $U^R$ is the generating functional of the connected part of $S$
and CF $G$ defined in (\ref{G(k)}) fulfills the role of the propagator.
Here and below the terminology from the quantum field theory, such as
the scattering matrix, the propagator, etc., has been used which may not
have much physical meaning in statistical models but the formalism we are
using is universal \cite{vasiliev1998} and the use of the established
terminology should facilitate comparison with physically different but
formally similar theories, such as the CPA and DMFT.

\begin{figure}
	\centering
	\includegraphics{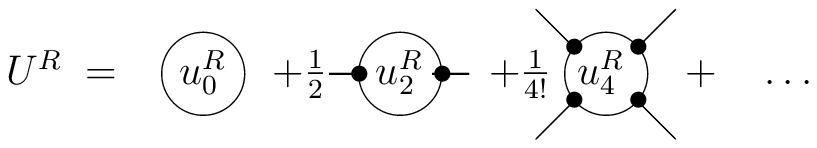}
	\caption{\label{fig1} Tree terms of expansion of 
	$U^R[{\boldsymbol\phi}]$ in powers of ${\boldsymbol\phi}$ for
	the functional even in the field variable.}
\end{figure}

The validity of expressions (\ref{Z-S}) and (\ref{lnZ-S}) is
easy to understand diagrammatically.  The connected diagrams for
S-matrix (see figure \ref{fig1}) differ from the diagrams for
$\ln Z$ in that they describe either (quasi)particle scattering or
reactions but not the free propagation. This is accounted for by the
first term on the r.h.s.\ of (\ref{lnZ-S}). Another distinctions
is that the diagrams in the generating functional of S-matrix
\cite{gamma_exp,tokar_new_1997,tan_topologically_2011,arXiv19}
\begin{equation}
S[{\boldsymbol\phi}]=\det(2\pi \hat{G})^{n/2}
	\exp\left(\frac{1}{2}{\partial_{\phi_\sigma}}
	\hat{G}{\partial_{\phi_\sigma}}\right)
	e^{\frac{r}{2}{\boldsymbol\phi}_i\cdot{\boldsymbol\phi}_i
	-H_I[{\boldsymbol\phi}]}
\label{S}
\end{equation}
are ``amputated'', that is, the tails in figure \ref{fig1}
correspond to field ${\boldsymbol\phi}$ and do not contain the factors
$\hat{G}$ that are present in the corresponding diagrams in $Z[{\bf
h}]$. This is remedied by replacing ${\boldsymbol\phi}$ in (\ref{S})
by ${\bf h}\hat{G}$ in (\ref{Z-S}) and (\ref{lnZ-S}). Note that in the
rightmost exponential an arbitrary quadratic term proportional to $r$
has been introduced  because of the arbitrariness in the definition of
the interaction Hamiltonian as discussed in the previous section. It is
compensated by $r$ in $G$ (\ref{G(k)}) so that $Z[{\bf h}]$ in (\ref{Z})
does not depend on this parameter, as can be easily shown formally
\cite{gamma_exp,tokar_new_1997,tan_topologically_2011,arXiv19}.

One of the reasons for introducing the generating functional of S-matrix
into the statistical theory was that the diagrams with $l$ tails in
the generating functional $\ln Z[{\bf h}]$ include the product of $l$
functions $G({\bf k})$.  In the critical region where all ${\bf k}$
are small this factor strongly depends on the momenta which makes
numerical approximations difficult. In contrast, the diagrams in
$U^R[{\boldsymbol\phi}]$ are devoid of these factors and the amputated
contributions admit approximation of the expansion coefficients by ${
\bf k}$-independent constants, as, e.g., in CPA and DMFT. This property
will be used below to introduce the LPA.

More important reason for using S-matrix is that it makes straightforward
generalization of the effective medium approaches of the CPA and DMFT type
to other cases, in particular, to the description of phase transitions
\cite{gamma_exp,tokar_new_1997,tan_topologically_2011,arXiv19}.
Formally the generalization is based on the exact expression for $G^R$
obtained by substitution of (\ref{lnZ-S}) in (\ref{ss_av}):
\begin{equation}
	G^R_{ij}=G_{ij}-G_{ii^\prime}(u^R_2)_{i^\prime j^\prime}G_{j^\prime j}.
	\label{def-G}
\end{equation}
This expression differs from (\ref{G(k)}) by the second term on the
r.h.s.\ so for self-consistency of the theory it should vanish which
means $u_2^R=0$.  From figure \ref{fig1} it is seen that without
the second diagram individual excitations can be perturbed only by other
excitations but not by the medium within which they propagate. In this
sense the medium can be called ``effective.

With constant $r$ the effective medium condition can be satisfied
only approximately.  In the CPA and DMFT the approximation consists
in assuming that the scattering is localized on the sites so that
$(u^R_2)_{ij}\propto\delta_{ij}$ and $r$ (the coherent potential or
the self-energy) is adjusted so that $(u_2^R)_{i=j}=0$.  The SSA,
however, gives a very poor description in the vicinity of the critical
point \cite{tokar_new_1997} that is why in the previous section the
condition (\ref{r=rR}) has been suggested.  As is easily seen by Fourier
transforming (\ref{def-G}) it amounts to
\begin{equation}
	u_2^R({\bf k}=0)=0.
	\label{U_2=0}
\end{equation}
Implementation of this condition requires explicit calculation of the
long-wave-length limit of $u_2^R$ which will be done below within
a lattice generalization of the functional RG approach developed in
\cite{1984} for $n$-vector models in the continuous space.
\section{The layer-cake renormalization scheme}
As was pointed out earlier, the generating functional of S-matrix is
convenient for approximate calculations so the RG equation will
be derived with the use of (\ref{S}) by casting it in the form
\begin{equation}
	e^{-U^R[{\boldsymbol \phi}]}=
	\exp
\left(\frac{1}{2}\sum_{\bf k}\partial_{{\phi}_{\sigma,{\bf k}}}
G({\bf k})\partial_{{\phi}_{\sigma,{\bf-k}}}
\right)e^{-U^0[{\boldsymbol \phi}]}
	\label{URU0}
\end{equation}
where
\begin{equation}
	{\phi}_{\sigma,{\bf k}}
=N^{-1/2}\sum_je^{-\mathrm{i}j\cdot{\bf k}}{\phi}_{\sigma j}.
	\label{phi-k}
\end{equation}
and
\begin{equation}
	U^0[{\boldsymbol\phi}] 
	=-\frac{r}{2}{\boldsymbol\phi}_i\cdot{\boldsymbol\phi}_i
	+H_I[{\boldsymbol\phi}]+\mbox{(f.i.t.)}.
	\label{U0}
\end{equation}
Here and everywhere below (f.i.t.) stands for field-independent terms
which will not be needed in calculations below. They, however, contribute
to the absolute value of the free energy and should be included in the
calculations if this quantity is of interest.

\begin{figure}
	\centering
	\includegraphics[width=0.85\textwidth,keepaspectratio]{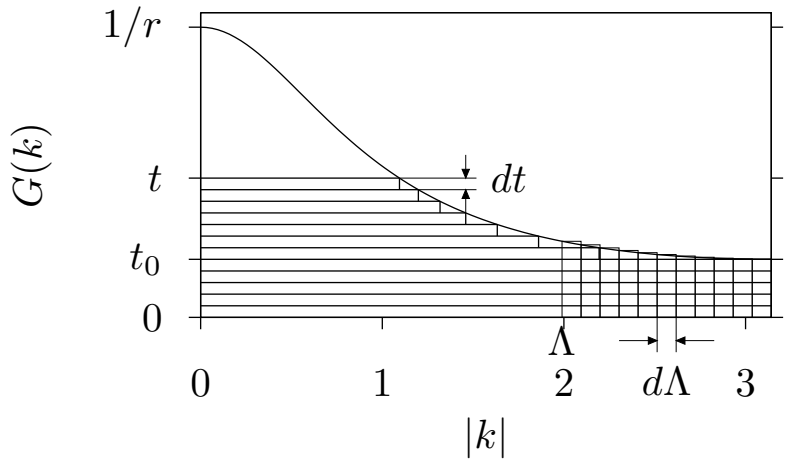}
	\caption{\label{fig2}Schematic illustration of the difference
	between the infinitesimal elimination steps in the layer-cake
	(horizontal slices) and the Wilsonian (vertical slices)
	renormalization schemes shown for 1D propagator (\ref{G(k)}).}
\end{figure}

Fully renormalized $U^R$ will be approximately calculated within the
renormalization scheme developed in \cite{1984} for continuous space and
adopted to the lattice case with the help of the layer-cake representation
of the propagator $G$ in (\ref{URU0}).  The layer-cake representation
is defined only for non-negative functions \cite{lieb2001analysis},
so we will assume that $G({\bf k})>0$ which is valid for all models
discussed below.  The representation is introduced by the identity
\begin{equation}
	G({\bf k})=\int_0^{G({\bf k})}dt^\prime
	=\int_0^{t_{end}} \theta[G({\bf k})-t^\prime]dt^\prime
	\label{layer-cake}
\end{equation}
where $t_{end}=1/r$ is the maximum value of the evolution parameter $t$
(see figure \ref{fig2}).

Integration over $t^\prime$ in (\ref{layer-cake}) effectively
splits the differential operator in (\ref{URU0}) into infinitesimal
contributions that can be used in the incremental renormalization
procedure \cite{wilson,1984}.  In the layer-cake renormalization scheme
the field components are integrated out layerwise \cite{1984}, as shown
in figure \ref{fig2}. Partly renormalized $U[{\boldsymbol\phi},t]$
at intermediate ``times'' is defined by the equation
\begin{equation}
	e^{-U[{\boldsymbol \phi},t]}=	\exp
\left(\frac{1}{2}\sum_{\bf k}
\int_0^{t} \theta[G({\bf k})-t^\prime]dt^\prime 
\partial_{{\phi}_{\sigma,{\bf k}}}
\partial_{{\phi}_{\sigma,{\bf-k}}}
\right)e^{-U^0[{\boldsymbol \phi}]}
	\label{U(t)}
\end{equation}
so satisfies the functional-differential evolution equation
\begin{equation}
	U^\prime_t=
	\frac{1}{2}\sum_{\bf k}\theta[G({\bf k})-t]
\left(	\partial_{{\phi}_{\sigma,{\bf k}}}
	\partial_{{\phi}_{\sigma,{\bf-k}}}U
	-\partial_{{\phi}_{\sigma,{\bf k}}}U
	\partial_{{\phi}_{\sigma,{\bf-k}}}U\right).
	\label{RG1}
\end{equation}
with the initial condition $U[{\boldsymbol \phi},t=0]=U^0[{\boldsymbol
\phi}]$ in (\ref{U0}). Equation (\ref{RG1}) has a form of the exact
RG equations derived, e.g., in \cite{wilson,nicoll_exact_1976}
(see bibliography to more recent literature in
\cite{BAGNULS200191,berges_non-perturbative_2002,RG2002review,latticeRG2010})
from which it differs only by the choice of the cut-off function and
by the absence of the rescaling of variables often used to obtain the
equations in scaling form.  The rescaling of (\ref{RG1}) could be easily
done but it would introduce into the equation the largest critical
exponent $d$ which simply reflects the fact that the free energy grows
with the linear system size $L$ as $L^d$.  Being the largest Lyapunov
exponent of the equation, however, it would noticeably deteriorate its
numerical solvability.

Equation (\ref{RG1}) is exact and its solution could be used to obtain
all CFs by differentiating (\ref{Z-S}) with respect to the source
field. However, its exact solution is impossible for nontrivial models
so approximations and simplifications are in order. One simplification
is achieved by restricting consideration to only homogeneous source
field ${\bf h}$ which in the momentum space will have only one non-zero
Fourier component ${\bf h}_{\bf k=0}$. Thus, to calculate magnetization in
(\ref{m}), only $U^R[{\boldsymbol\phi}_{\bf k=0}]$ needs to be kept in
(\ref{lnZ-S}). In view of this we will assume that all field components
in (\ref{RG1}) with the momenta outside the ``Fermi surface'' defined by
the condition $G({\bf k})=t$ or
\begin{equation}
	\epsilon({\bf k})=t^{-1}-r\equiv E_F	
	\label{E_F}
\end{equation}
will be set equal to zero so at $t=t_{end}$ only the component
${\boldsymbol\phi}_{\bf k=0}$ will survive. In this case the functional
$U^R$ effectively becomes a function of a single $n$-vector variable.
\subsection{The LPA}
As was noted in the Introduction, the LPA is similar to the such
approximations as CPA and DMFT in that currently it cannot be rigorously
justified in the strong coupling case. (In the limit of weak coupling all
approximations reproduce correctly the lowest-order perturbative terms so,
on the one hand, the perturbation theory justifies them in this limit,
on the other hand, makes them superfluous.) The gradient expansion
\cite{golner,IVANCHENKO1990100} frequently invoked to substantiate LPA
is inapplicable in lattice models, so only heuristic arguments, such
as those presented below, can be given to qualitatively understand, at
least qualitatively, why LPA gives reasonable quantitative approximation
to for such models, as will be illustrated by numerical calculations in
section \ref{numerical}.

Thus, designating LPA as an approximation is somewhat misleading in the
lattice case where it is rather an {\em ansatz} which consists in assuming
that in models with the local initial $U^0$ (\ref{U0}), $U[{\boldsymbol
\phi},t]$ will preserve the locality throughout the evolution described
by the RG equation (\ref{RG1}).  The simplification achieved consists in
that the functional of $N\times n$-dimensional lattice field can be fully
characterized by a function $u({\bf x},t)$ of one $n$-dimensional vector
${\bf x}$ which in RG literature is usually called the local potential. To
derive RG equation for $u$ we first note that in Fourier representation
(written here for simplicity for $n=1$)
\begin{eqnarray}
	&&U[{\phi},t]=N\sum_{l=0}^\infty\sum_{\{\bf{k}_m\}}N^{-l/2}u_l(t)
\delta\left(\sum_{m=1}^l{\bf k}_m\right)\frac{1}{l!} 
\prod_{m=1}^l\phi_{{\bf k}_m}\\
&&\to u(x,t)=\sum_{l=0}^\infty u_l(t)\frac{x^l}{l!} \nonumber
	\label{U(k)}
\end{eqnarray}
(where $\delta$'s are the lattice Fourier transform of the Kronecker
delta) the locality means that the expansion coefficients $u_l(t)$
do not depend on the field momenta ${\bf k}_m$.

In this approximation the action of the second derivatives on the r.h.s.\
of (\ref{RG1}) on each term in (\ref{U(k)}) amounts to the elimination 
of two field variables which momenta ${\bf k}$ and $-{\bf k}$ cancel out
in the delta-function argument so all dependence on ${\bf k}$ separates
into the factor common to all terms in the expansion
\begin{eqnarray}
	p(t)&=&\frac{1}{N}\sum_{\bf k}\theta\left[G({\bf k})-t\right]
	=\frac{1}{N}\sum_{\bf k}\theta[t^{-1}-r-\epsilon({\bf k})]\nonumber\\
	&=&D_{tot}(t^{-1}-r)=\int_0^{t^{-1}-r}dE D(E).
	\label{p}
\end{eqnarray}
Here the second equality on the first line follows from the fact that the
value of theta-function does not change when its argument is multiplied
by the positive function $[\epsilon({\bf k})+r]/t$ and the first equality
on the second line follows from the observation that $\theta$-function
is a $T\to0$ limit of the Fermi-function and thus the summation over
momentum counts the total number of states below the ``Fermi energy''
$E_F$ (\ref{E_F}) in the band with dispersion $\epsilon({\bf k})$
and so is equal to the integrated density of states $D(E)$ at $E_F$
\cite{latticeRG2010,arXiv19}.  This makes possible to write down the
first term in (\ref{RG1}) in terms of function $u$ as
\begin{equation}
	\frac{1}{2}\sum_{\bf k}\theta[G({\bf k})-t]
\partial_{{\phi}_{\sigma,{\bf k}}}
\partial_{{\phi}_{\sigma,{\bf-k}}}U\rightarrow 
\frac{1}{2}p(t)\nabla^2u({\bf x},t). 
	\label{1st-term}
\end{equation}

The second term on the r.h.s.\ of (\ref{RG1}) is more convenient to
analyze in the lattice coordinates. Substituting $U$ in the LPA from
(\ref{U(k)}) into the second term one gets
\begin{eqnarray}
-\frac{1}{2}\sum_{\bf k}\theta[G({\bf k})-t]
	\partial_{{\phi}_{\sigma,{\bf k}}}U
	\partial_{{\phi}_{\sigma,{\bf-k}}}U
	&=&-\frac{1}{2}U_{{\boldsymbol\phi}_{\sigma,i}}  \Delta_{ij}
U_{{\boldsymbol\phi}_{\sigma,j}}\nonumber\\
&\leadsto& -\frac{1}{2}[\nabla u({\bf x},t)]^2 
	\label{2nd-term}
\end{eqnarray}
where
\begin{equation}
	\Delta_{ij}=\sum_{\bf k}\theta[G({\bf k})-t]
	e^{\mathrm{i}{\bf k}\cdot(i-j)}
	\label{Delta}
\end{equation}
and the wiggly arrow means an approximate mapping, as explained below.  If
$\theta$ in (\ref{Delta}) is equal to unity in the whole BZ, $\Delta_{ij}$
becomes equal to the conventional Kronecker delta and the term on
the r.h.s.\ becomes strictly local and LPA is fully justified. This
special case will be considered in the next subsection.  However, as $t$
evolves toward $t_{end}$, $\Delta_{ij}$ becomes progressively more and
more spatially extended with characteristic half width $O(1/k_c)$ where
$k_c$ is the maximum absolute value of the cutoff momenta at the Fermi
surface (\ref{E_F}). On the other hand, $\Delta_{ij}$ remains peaked at
$i=j$, it is symmetric in $i$ and $j$, and is normalized to unity:
$\sum_{i({\rm or\ }j} \Delta_{ij}=1$.  Thus, it can be considered as
a smeared approximation to the Kronecker delta which can be used to
justify the approximation of the term on the r.h.s.\ of (\ref{2nd-term})
by the squares of the local derivatives.  Of course, from the mathematical
standpoint the approximation severely worsens when $k_c\to0$ as $t\to
t_{end}$. But we remind that in the process of elimination of high
momentum components the nonzero field components ${\boldsymbol\phi}_{\bf
k}$ also are restricted to the same momentum range $|{\boldsymbol k}|\le
k_c$ so in the lattice coordinates ${\boldsymbol\phi}_{\sigma,i}$ as well
as $U({\boldsymbol\phi}_{\sigma,i})$ entering (\ref{2nd-term}) are slowly
varying functions of the lattice coordinates on the same scale $O(1/k_c)$.
Thus, approximation $\Delta_{ij}\approx \delta_{ij}$ in (\ref{2nd-term})
may be quite consistent in this particular case. Besides, the gradient
expansion \cite{golner,IVANCHENKO1990100} which in the momentum space
is equivalent to expansion in powers of ${\bf k}_m$ may also be invoked
to substantiate LPA in the region of small momenta.

Substituting (\ref{1st-term}) and (\ref{2nd-term}) in (\ref{RG1}) one
arrives at the equation 
\begin{equation}
	u_t = \frac{1}{2}\left[p(t)\nabla_{\bf x}^2 u 
	- (\nabla_{\bf x} u)^2\right].
	\label{LPA}
\end{equation}
which is the lattice generalization of the LPA equation derived in
\cite{1984}.
\subsection{\label{partial}Exact partial renormalization}
Differential equations of the form (\ref{LPA}) currently cannot be
solved analytically for arbitrary $p(t)$, so numerical solutions should
be used.  However, even numerical treatment meets with difficulties in
some cases. For example, in the case of the {\em spin} models when the
initial Hamiltonian (\ref{delta}) is singular and is not differentiable
numerically. Fortunately, when $p(t)=Const$ equation (\ref{LPA}) is
exactly solvable which can be used to regularize the initial conditions
as follows.

From the definition of $p$ (\ref{p}) it can be seen that for sufficiently
small $t$ the ``Fermi energy'' $t^{-1}-r$ will exceed the width of the
quasiparticle band that extends from zero to $\max_{\bf k}\epsilon({\bf
k})$ the integrated densities of states (DOS) will be equal to its
maximum unit value when $t$ satisfies
\begin{equation}
	0\le t\le t_0=\min_{\bf k}G({\bf k})
	=\frac{1}{\max_{\bf k}\epsilon({\bf k})+r}
	\label{t0}
\end{equation}
(see figure \ref{fig2}). In this range equation (\ref{LPA}) with
$p=1$ by substitution $e^{-u}$ can be transformed to the $n$-dimensional
linear diffusion equation which solution with the use of the diffusion
kernel reads
\begin{equation}
	e^{-u({\bf x},t_0)}
=\frac{1}{(2\pi t_0)^{n/2}}\int d{\bf x}_0\,
\exp{\left(-\frac{({\bf x-x}_0)^2}{2t_0}\right)}e^{-u^0{\bf x}}
\label{f_ini}
\end{equation}
where $u^0$ is the potential corresponding to (\ref{U0}).

Thus, in the case of local Hamiltonians the RG equations,---both, exact
and approximate,---can be integrated from $t_0$ to $t_{end}=1/r$ with
the initial condition (\ref{f_ini}) (for the exact equation the inverse
map $u({\bf x},t_0)\to U[{\boldsymbol\phi},t_0]$ should be performed).
It is pertinent to note here that when $r$ becomes large, the integration
span $\Delta t=t_{end}-t_0$ shrinks as $\sim1/r^2$. Function $G({\bf k})$ 
flattens and a good approximation can be obtained by neglecting its 
${\bf k}$-dependence and approximating $G({\bf k})$ by its average
\begin{equation}
	g=N^{-1}\sum_{\bf k}G({\bf k}).
	\label{g}
\end{equation}
In this case integration over $t$ in (\ref{RG1}) and (\ref{LPA}) is
not needed anymore because renormalized $U^R$ and $u^R$ are given by
(\ref{f_ini}) with $t_0=g$. Augmenting this with the self-consistency
condition $(u^R_2)_{i=j}=0$ (see (\ref{def-G})) one arrives at the SSA
such as the CPA \cite{gamma_exp,tokar_new_1997}. 
In the case of large $r$ it is straightforward to calculate corrections
to the SSA.  Formally (\ref{URU0}) can be seen as a closed form of
the perturbative expansion so by choosing in (\ref{f_ini}) $t_0=g$ and
simultaneously subtracting $g$ from $G$ in (\ref{URU0}) one arrives at
the expansion with the renormalized local potential (\ref{f_ini}) and the
propagator \cite{gamma_exp,tokar_new_1997}.
\begin{equation}
	\tilde{G}({\bf k}) = {G}({\bf k})-g.
	\label{G-g}
\end{equation}
When $r$ is large the deviations of $G{\bf k})$ from $g$ will be small
(see figure \ref{fig2}) so $\tilde{G}$ can be used as an expansion
parameter.  The Feynman diagrams in this expansion would simplify
because the tadpole contributions will be implicitly accounted for by
the redefined propagator (\ref{G-g}) which is similar to the normal
ordering in quantum many-body theory. So the relative importance of
the correction terms will be defined by the number of propagator lines
$\tilde{G}({\bf k})$ in the corresponding diagrams.
\section{\label{numerical}Numerical results}
The LPA RG equation (\ref{LPA}) could be readily integrated in
the symmetric phase but below the critical temperature $T_c$ the
solution exhibited singular behavior due to the physics of the
phase coexistence region.  In the $n=1$ case this can be easily
understood as follows (see discussion of this point in the RG context
in \cite{maxwell_construction,caillol_non-perturbative_2012}). In the
zero external field the statistical ensemble below $T_c$ consists of
a mixture of two pure phases with saturated magnetizations $\pm m_0$
so in the coexistence region the magnetization may take any value in the
interval $m\in[-m_0,m_0]$. But an infinitesimal external field $h=0^\pm$
will bring the magnetization either to $m_0$ or to $-m_0$ which in
the case $m\not=\pm m_0$ means infinite susceptibility.  However,
if the magnetization is saturated and is equal, e.g., to $m_0>0$
and the infinitesimal field also is positive the susceptibility will
remain finite which means discontinuity at $h=0$ because for $h=0^-$
magnetization will jump to the value $-m_0$.

According to (\ref{s_av}), in the LPA the magnetization in the homogeneous
(ferromagnetic) Ising model ($n=1$) is
\begin{equation} 
	m = \frac{1}{N}\frac{\partial\ln Z(h)}{\partial h} = x-u_x^R/r
	\label{m} 
\end{equation} 
where $x=h/r$, $u^R=u(x,t_{end})$ and in the last equality use has been
made of equations (\ref{Z-S}), (\ref{S}), (\ref{U(k)}) and (\ref{G(k)}). The 
susceptibility is found from (\ref{m}) as
\begin{equation}
	\chi = m_h = \frac{1}{r}- \frac{1}{r^2}u_{xx}^R.
	\label{chi}
\end{equation}
At a finite distance from $T_c$ the correlation length, hence, $1/r$
remain finite, so it is the second derivative in (\ref{chi}) that is
responsible for the infinite value of $\chi$.

The origin of the singularity can be understood from the particular solution
of equation (\ref{LPA}) 
\begin{equation} 
	u^{G}({\bf x},t)=\frac{{\bf x}^2}{2(t-t_{end})}+\mbox{(f.i.t.)}.
	\label{u_G}
\end{equation} 
It formally corresponds to the Gaussian model but is non-physical because
$t$ is always smaller than $t_{end}$ so the Hamiltonian corresponding
to (\ref{u_G}) is not bounded from below. However, it may be useful to
qualitatively understand the singularity in the solution for physical
models because below $T_c$ all $u^0({\bf x})$ (see, e.g., \ref{ini})
have negative curvature near ${\bf x}=0$ and their solutions may develop
behavior similar to (\ref{u_G}) (above $T_c$ the negative curvature does
not survive the renormalization and becomes positive at $t=t_{end}$).

The unbounded derivatives in ${\bf x}$ and $t$ indeed arose in the
numerical solutions of the LPA equation (\ref{LPA}) which made it
difficult to deal with numerically below $T_c$. Fortunately, the
singularities disappear under the Legendre transform given by equations
(\ref{y-x}) and (\ref{v-u}) which are a $t$-dependent generalization of
the transform suggested in \cite{x-yLegendre}. Under the transform
the Gaussian solution (\ref{u_G}) takes the form
\begin{equation} 
	v^{G}({\bf y},t)=-\frac{{\bf y}^2}{2\Delta t}+\mbox{(f.i.t.)},
	\label{v_G}
\end{equation} 
in which all terms are bounded.

Though in the symmetric phase equation (\ref{LPA}) can be integrated
for any $n$, it showed lesser stability near the critical point than
the transformed equations, presumably because of the closeness to the
ordered state. Therefore, in the calculations below the transformed
equations will be used both above and below $T_c$.
\subsection{The symmetric phase}
In the symmetric phase the transformed equation derived from (\ref{LPAv}),
(\ref{q}) and (\ref{u_xixi}) reads
\begin{equation}
	w_t=\frac{p(t)}{2}\left(\frac{(n-1)w_q}{1+(t-t_0)w_q}
+\frac{w_q+2qw_{qq}}{1+(t-t_0)(w_q +2qw_{qq})}\right)
	\label{RGw}
\end{equation}
where the arbitrary constant $c$ was chosen to be equal to $t_0$. 
With this choice the initial condition is easily found from 
(\ref{y-x}), (\ref{v-u}) and (\ref{q}) as
\begin{equation}
	w(q,t_0)=u(\sqrt{2q},t_0)
	\label{w-ini}
\end{equation}
with $u({\bf x},t_0)$ calculated in \ref{ini}; here the superscript
``$(n)$'' was omitted for consistency with (\ref{RGw}) which holds for
all $n\ge1$.  

The self-consistency condition (\ref{U_2=0}) in the fully symmetric
case means that all derivatives $u_{x_\sigma x_{\sigma^\prime}}$
in (\ref{uxx-vyy}) vanish so all $v_{y_{\sigma}y_{\sigma^\prime}}$
are also equal to zero. In the $O(n)$-symmetric case this leads via
(\ref{v_yiyj}) to
\begin{equation}
	w_q^R|_{q=0}=0.
	\label{scc1}
\end{equation}

Equation (\ref{RGw}) and (\ref{scc1}) with the initial condition
(\ref{w-ini}) has been solved numerically for the $n$-vector spin models
with
\begin{equation}
	\epsilon({\bf k})=K\sum_{i_{NN}}(1-e^{\mathrm{i}{\bf k}\cdot i_{NN}})
	\label{J-NN}
\end{equation}
where $K=|J|/k_BT$ ($J$ the ferromagnetic coupling between NN spins)
and $\{i_{NN}\}$ are the lattice vectors connecting the site at the origin
with all NN sites.

In the calculations $D_{tot}(E)$ for dispersions (\ref{J-NN}) were
obtained with the use of the expressions derived in \cite{Jelitto1969609}
for the DOS of SC, BCC and FCC lattices.  For general dispersion
straightforward numerical integration of (\ref{p}) by the Monkhorst-Pack
method \cite{special_points} can be efficient because the integrated
DOS $D_{tot}(E)$ is much less structured than the DOS itself. In
case of necessity, the region of small $|{\bf k}|$ near the bottom
of the band where $\epsilon({\bf k})\propto {\bf k}^2$ can be treated
analytically. Thus, application of the LPA to the pair spin interactions
of arbitrary range \cite{ducastelle,Zunger2004,us-ordering-potentials}
should cause no problems.

The LPA equations have been solved by the method of lines with the use
of the Fortran LSODE routine \cite{lsode}.  The number of equations
used varied in the range 2-4 thousands until convergence has been
reached. Qualitative behavior of the solutions both in the symmetric
and in the ordered phases agreed with the results of previous studies
\cite{maxwell_construction,caillol_non-perturbative_2012,arXiv19}.
Near the critical temperature $t_{end}=1/r\to\infty$ but in numerical
integration the interval is necessarily bounded so $K_c$ was found
by extrapolating several $r$ calculated at $K\gtrsim K_c$  to $r=0$
according to the scaling relation
\begin{equation}
	r=C_\pm^{-1}\tau^{2\nu}=C_\pm^{-1}\tau^{\gamma} 
	\label{r(tau)}
\end{equation}
where $\tau = |1-K_c/K|$ and $\gamma=2\nu$ because in the LPA $\eta=0$.
The correlation length in the Ising model both above (+) and below (-)
$T_c$ has been estimated on the basis of the asymptotic behavior of $G$
in real space which can be found by the inverse Fourier transform of
(\ref{G(k)})
\begin{equation}
\xi = \sqrt{K/r}\stackrel{T\to T_c}{\simeq} f_{\pm}\tau^{-\nu}. \label{f+} 
\end{equation}
For consistency, the critical exponents $\nu$ were found from the scaled
form of the RG equation (\ref{RGw}) as explained in \cite{1984}. The
obtained values of $\nu$ were equal to 0.65 for $n=1$, 0.71 for $n=2$
and 0.76 for $n=3$ and were similar but larger than those systematized
in table 2 in \cite{berges_non-perturbative_2002}. They were closest to
the values of $\nu$ calculated in \cite{berges_non-perturbative_2002}
by being larger on 0.01. This apparently is a consequence of a similar
non-perturbative RG approach used by the authors. The difference
caused by the difference in $\eta$ which in the calculations of
\cite{berges_non-perturbative_2002} was not equal to zero but obtained
from RG equations.

The calculated values of $K_c$ at the critical point found with the use
of equations (\ref{RGw}) are presented in table \ref{T_c}.
\begin{table}
\caption{\label{T_c}Dimensionless inverse critical temperatures of the
$n$-vector spin models on cubic lattices calculated in the LPA. The
errors have been estimated by comparison with data from MC simulations
\cite{bcc-fcc-diamond-Kc} for $n=1$ (the Ising model) and with the high
temperature expansion for $n=2$ and 3 \cite{n-vector-models}.}
\begin{tabular}{@{}llcl}
\hline
$n$&Lattice&$K_c$&Error\\
\hline
1&FCC&0.1023&0.2\%\\
1&BCC&0.1579&0.3\%\\
1&SC &0.2235&0.8\%\\
2&BCC&0.3225&0.6\%\\
2&SC &0.4597&1.2\%\\
3&BCC&0.4905&0.8\%\\
3&SC &0.7025&1.4\%\\
\hline
\end{tabular}
\end{table}
\subsection{\label{order}Ordering in the Ising model}
According to (\ref{s_av}), to calculate the spontaneous magnetization in
zero external field we still need to take into account the source field
$h$.  In the $n=1$ case the Legendre transform (\ref{x-y}) simplifies to
\begin{eqnarray}
	\label{y-x1}
y&=&x-(t-t_0)u_{x}\\
	\label{v-u1}
	v&=&u-\frac{1}{2}(t-t_0)u_{x}^2
\end{eqnarray}
where the arguments of $u(x,t)$ and $v(y,t)$ have been omitted for
brevity and constant $c$ was chosen to be equal to $t_0$ because with
this choice there is no need to perform numerical Legendre transform
of the initial condition (\ref{n=1}) needed in some renormalization
schemes \cite{latticeRG2010}:
\begin{equation}
	v(y,t_0)=u^{(1)}(x,t_0)|_{x=y}.
	\label{v0=u0}
\end{equation}
The RG equation in this case reads (see \ref{x-y})
\begin{equation}
v_t = \frac{p(t)v_{yy}}{2[1+(t-t_0)v_{yy}]}.  \label{the_eq2}
\end{equation}
The dependence of $v^R$ on $h$ at the end of renormalization is defined
parametrically with the use of the expression for $h$
\begin{equation}
x = h/r = y+\Delta t v_y^R	
	\label{x(y)}
\end{equation}
which follows from (\ref{y-x1}) and (\ref{v=u_diff}) with $t=t_{end}$.
The unknown parameter $r=1/t_{end}$ in (\ref{x(y)})
is  fixed by the self-consistency condition (\ref{U_2=0}) which according
to (\ref{uxx-vyy}) and (\ref{u_1-v_1}) now reads
\begin{equation}
	v^R_{yy}|_{h=0^+}=0
\label{v_yy=0}
\end{equation}
where $h$ should be expressed through $y$ according to (\ref{x(y)}).
At $h=0$ there always exists a trivial solution $y=0$ due to the
symmetry. But below $T_c$ two stable solutions with $y=\pm y_0\not=0$
appear which correspond to the states with spontaneous magnetization
\begin{equation}
	m_0 = y_0-t_0v^R_y|_{y_0}
	\label{m0}
\end{equation}
as can be found from (\ref{m}), (\ref{x(y)}) and (\ref{v=u_diff}).

Numerically found solution of equation (\ref{the_eq2}) with the initial
condition (\ref{v0=u0}) and the self-consistency condition (\ref{v_yy=0})
exhibited the same qualitative behavior as found previously in
\cite{caillol_non-perturbative_2012}. Namely, the inverse susceptibility
was equal to zero in the interval $(0,m_0)$ (the solution for negative
$m$ can be obtained by symmetry) and jumped to $\chi^{-1}=r$, at $m_0$,
in accordance with (\ref{chi-1}). In particular, it was found that the
Gaussian solution (\ref{v_G}) describes the solution for the Ising model
in the ordered phase so accurately that the precision of the calculations
was insufficient to see the difference.  This can be partly explained
by the fact that the $y$-dependent part of (\ref{v_G}) is stationary and
from (\ref{the_eq2}) it can be seen that small $y$-dependent deviations
from $v^G$ do not grow when $t\to t_{end}$ so (\ref{v_G}) might have
been a fixed point of the LPA RG equation if $t\to\infty$. However,
in the SC-LPA the largest $t$ is equal to $t_{end}=1/r$ which away from
the critical point is finite. So the close proximity of the solution to
the Gaussian model seen in the calculations might be due to relatively
large values of $t_{end}$ in our calculations performed in the region
within 10\% distance from $T_c$.

Similarly, it was impossible to establish numerically whether
the discontinuity in the inverse susceptibility is genuine or
is just a very steep continuous transition (see discussion in
\cite{caillol_non-perturbative_2012}). But anyway the LPA is not
exact, so taking into account the narrowness of the finite difference
step in $\Delta y=O(10^{-3})$ within which the inverse susceptibility
changed from zero to $r$ and the fact that the exact behavior is well
understood qualitatively, the approximately correct behavior obtained
numerically can be fitted to the exact one with the use of an analogue
of the Maxwell construction appropriately adopted to this case (cf.\
\cite{maxwell_construction}).

In the calculations the value of $m_0$ has been calculated from (\ref{m0})
at the point $y_0$ where both conditions (\ref{x(y)}) with $h^+=0$
and (\ref{v_yy=0}) has been satisfied by appropriate choice of $r$.
The possibility to satisfy two conditions with one adjustable parameter
was due to the fact that the jump occurs only in $v^R_{yy}$ but $v^R_y$ is
continuous. And because the Ising model solution for $y\le y0$ coincided
with the Gaussian solution (\ref{v_G}), $h$ in (\ref{x(y)}) was equal
to zero everywhere in the interval $[0,y_0]$ including the endpoint,
so effectively only the self-consistency condition (\ref{v_yy=0}) had
to be satisfied.

\begin{figure}
	\centering
\includegraphics[viewport = 0 0 283 200,scale=0.7,keepaspectratio]{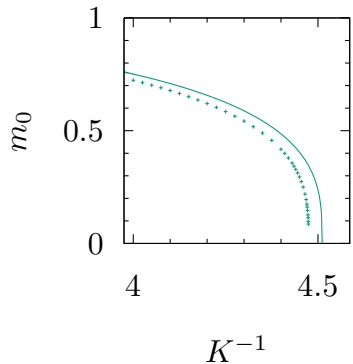}
	\caption{\label{fig3}Magnetization curve near the critical
	point as obtained in the LPA (symbols) and the fitting curve
	(\ref{talapov}) to the exact MC simulations with the parameters
	from \cite{talapov_M(t)} (solid line).}
\end{figure}

The magnetization curve obtained by the above procedure is shown in
figure \ref{fig3} together with the exact MC simulations data from
\cite{talapov_M(t)} accurately described by the expression
\begin{equation}
m_0(\tau) = \tau^\beta(a_0-a_1\tau^{\theta_W}-a_2\tau)
	\label{talapov}
\end{equation}
This expression has been also fitted to LPA curve but with $\beta$
equal to $\beta_{LPA}=\nu/2$ and the Wegner exponent $\theta_W$ set to
0.5 \cite{talapov_M(t)}.  In table \ref{amplitudes} the parameters of both
fits are compared.  As is seen, the LPA parameters $a_{1-2}$ deviate quite
appreciably from the (rounded) exact values of \cite{talapov_M(t)} but it
has to be born in mind that the parameters refer to the correction terms
that enter (\ref{talapov}) multiplied by positive powers of $\tau$. The
latter was smaller than 0.1 in the calculations so, on the one hand, the
contribution of the terms to the magnetization curve was reduced by these
factors, on the other hand, their fit for the same reason was not very
accurate and might improve with improved precision of the calculations.
\begin{table}
\caption{\label{amplitudes}The amplitudes entering the scaling relations
(\ref{r(tau)}), (\ref{f+}) and (\ref{talapov}) calculated in the
present work compared with the high temperature expansions data of
\cite{liu_fisher89} ($C_\pm,f_\pm$) for $\gamma=1.25$ which is closest
to the LPA value 1.3 and with the MC simulations of \cite{talapov_M(t)}
($a_0-a_2$).}
\begin{tabular}{@{}lccccccc}
\hline
&$C_+$&$f_+$&$C_-$&$f_-$&$a_0$&$a_1$&$a_2$\\
\hline
LPA&1.06&0.487&0.21&0.22&1.62&0.23&0.37\\
\cite{liu_fisher89,talapov_M(t)} &1.06&0.485&0.21&0.25&1.69&0.34&0.43\\
\hline
\end{tabular}
\end{table}

The temperature range in the simulations was restricted to 10\% distance
from $T_c$ for the following reasons. First, the correlation length
at the lowest temperature $\sim0.8$~l.u.\ was already smaller than the
lattice constant which means that the system was far from the critical
region which was of the main interest in the present study.  Second,
at lowering temperature the numerical integration considerably slowed
down so that convergence to the self-consistent solution by iterations
was difficult to achieve. It is quite possible, however, that the SC-LPA
can describe the low-temperature magnetization in an asymptotically
exact way. As $T\to0$, $r$ grows very fast so $G({\bf k})$ flattens
(see section \ref{partial}) and the SSA should be adequate. It does
reproduce correctly the leading asymptotic correction to the saturated
magnetization $m_0=1$ and the LPA may follow the suite.  However,
at low temperatures $r$ grows by the Arrhenius law $r\propto e^{AK}$
with $A=O(10)$ so the integration range shrinks as $1/r\sim e^{-AK}$
and at the same time the second derivative $u_{xx}^0$ in (\ref{n=1})
grows as $r^2$ which poses serious computational problems. Presumably,
an analytic expansion of the kind discussed in section \ref{partial} or
in \cite{gamma_exp} wold be more suitable in the low-temperature region.
\subsection{\label{beta-brass}Ordering in beta brass}
As is known, the Ising model on a bipartite lattice in zero external
field ordered antiferromagnetically can be mapped onto ferromagnetic
model by flipping the spins on every second site and simultaneously
changing signs of the spin interactions in such a way that the value
of the Hamiltonian remained unchanged. So the formalism developed for
the ferromagnetic case is fully applicable to the ordering of, e.g.,
an equiatomic BCC alloy, such as the beta brass \cite{beta-brass2016}.

The interpretation of the experimental data on ordering in this alloy
on the basis of approximate solutions of the NN Ising model showed an
overall good agreement with experiment (see \cite{beta-brass2016} and
references therein). However, theoretical value $\beta=0.313$ of the order
parameter exponent used by the authors was about 4\% smaller than the
value $\beta=0.3265$ predicted by the RG theory \cite{RG2002review} though
it fitted well experimental data in the $\gtrsim1\%$ vicinity of $T_C$.
In a broader fitting range $0<\tau<0.04$ the difference was even larger
($\gtrsim10\%$) \cite{beta-brass2016}.  The reason for the discrepancy
apparently lies in the inevitably finite temperature intervals near $T_c$
used in experimental measurements while the RG predictions are strictly
valid only infinitesimally close to the critical temperature. Thus,
explanation for the deviation of the experimental values from RG theory
should be sought in the influence of non-universal contributions.

\begin{figure}
	\begin{center}
	\includegraphics[keepaspectratio,scale = 0.8]{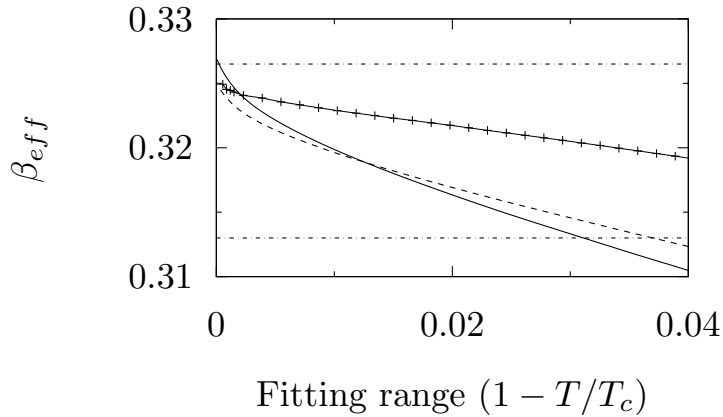}
	\end{center}
	\caption{\label{fig4} Symbols (the line is to guide
	the eye):  exponent $\beta$ as obtained by the fit of the
	scaling law $m_0\propto\tau^{\beta_{eff}}$ within the interval
	$0-\tau$ to the magnetization $m_0$ calculated with the use of the
	SC-LPA equations for the BCC Ising model; solid and dashed lines:
	same fit to equation (\ref{talapov}) with the parameters taken
	from the second and the first lines in table \ref{amplitudes},
	respectively; the upper horizontal dashed-dotted line is the
	best known universal value for the exponent \cite{RG2002review},
	the lower line is the value used in \cite{beta-brass2016}.}
\end{figure}

Indeed, by fitting expression (\ref{talapov}) to the scaling form with
an effective exponent $\beta_{eff}$ (see figure \ref{fig4}) one
finds noticeable deviation from the RG value, though not as large as
observed experimentally.

Though the LPA calculations shown in figure \ref{fig4} do not
agree with the experimental data on $\beta$ \cite{beta-brass2016}
quantitatively, the deviations from the RG value are of similar order
of magnitude.  Also, their absolute value grows with the widening of the
fitting range qualitatively similar to findings in \cite{beta-brass2016}
and are of the correct sign.  The latter point is quite nontrivial
because it is usually expected that when going away from the critical
point the system tends to exhibit the mean-field behavior \cite{wilson}
with $\beta_{MF}=0.5>\beta$. So {\em a priori} one would expect
$\beta_{eff}>\beta$ in contradiction to experiment but in accordance
with the LPA and with the exact MC simulations \cite{talapov_M(t)}.
Also noticeable is the fact that though the SC and BCC Ising models
differ in the critical region only by irrelevant variables, as is seen
from figure \ref{fig4}, the deviation of $\beta_{eff}$ from $\beta$
is quite different in the two cases which means that the influence of
irrelevant variables is strong. On the other hand, from the results of
the band structure theory of order in alloys it looks highly implausible
that the ordering potential would be accurately represented by the
NN Ising model \cite{ducastelle,Zunger2004}.  Much more likely that
the pair interactions would extend on many coordination spheres and
also multi-site cluster interactions may contribute to the ordering.
Therefore, the non-universal contributions can be very different from the
simple NN case used in the calculations shown in figure \ref{fig4}
and might be strong enough to bring the LPA values of $\beta_{eff}$
close to those found in \cite{beta-brass2016}.
\section{\label{discussion}Discussion}
Numerical calculations of the last section have shown that the
self-consistent RG equation derived in the present Letter make
possible calculation of non-universal quantities in $n$-vector spin
models within LPA in good agreement with the values known from reliable
techniques such as the high-temperature expansions and the Monte Carlo
simulations \cite{n-vector-models,talapov_M(t),bcc-fcc-diamond-Kc}.
The accuracy of calculation of the critical temperatures can be
sufficient to many practical purposes, e.g., in application to phase
equilibria in alloys where the errors in microscopic Hamiltonian
parameters determined either theoretically or experimentally
\cite{ducastelle,Zunger2004,us-ordering-potentials} are currently larger
than $0.2-0.3\%$ errors in the LPA values of $T_c$ for BCC and FCC Ising
models (see table \ref{T_c}).

The main deficiency of the SC-LPA approach is that, similarly to CPA
and DMFT, it is a closed-form approximation which cannot be rigorously
justified and/or improved in the strong coupling case (in the case of
weak coupling all effective medium theories are exact to the leading
order in the interaction and are normally superfluous). As was mentioned
in the Introduction, a natural way of going beyond the SSA such as
CPA and DMFT by analogy with which SC-LPA has been devised is to
resort to their cluster generalizations which can be systematically
improved and which are based on the same self-consistency condition
\cite{gamma_exp,tokar_new_1997,maier_quantum_2005} as the SSA thus
making generalization relatively straightforward.  Schematically this
can be done as follows.  First, in (\ref{G-g}) constant $g$ should be
replaced by a function of ${\bf k}$ consisting of a finite sum of the
lattice Fourier terms
\begin{equation}
	g({\bf k})=\sum_{|l|\leq L^c}g_le^{\mathrm{i}l\cdot{\bf k}}
	\label{g(k)}
\end{equation}
where $L_c$ is some spatial cut-off so in the lattice coordinated the
matrix elements ${g}_{ij}$ will vanish beyond the cluster of radius $L_c$:
$g_{ij}|_{|i-j|>L_c}=0$. Now, if the system is far from criticality the
matrix elements of the propagator $G$ will exponentially attenuate at
large separations  \cite{gamma_exp} and may be neglected for $|i-j|>L_c$
with the accuracy $O(e^{-L_c/\xi}$.  In such a case by choosing
$g_l\simeq G_{i-j=l}$ in (\ref{g(k)}) one may neglect $\tilde{G}$ in
(\ref{G-g}) so the partition function can be calculated by the cluster
techniques with the use of the clusters of radius $\simeq L_c$ (see
\cite{tokar_new_1997,tan_topologically_2011}).

As is easy to see, the approximation will break down in the critical
region. For example, at the critical point $G$ in (\ref{G(k)}) is
singular at ${\bf k}=0$ but no finite Fourier sum (\ref{g(k)}) will
reproduce the singularity. Still, one can choose $g_l$ in (\ref{g(k)})
in such a way that $g({\bf k})$ will satisfactorily approximate $G({\bf
k})$ everywhere in BZ with the exception of a region surrounding ${\bf
k}=0$ with $|{\bf k}|<k_c$, where the cut-off $k_c\sim1/L_c$.  Now the
contributions due to $g$ can be calculated within the cluster method
while the remaining singular part $\tilde{G}$ for $|{\bf k}| <k_c$
accounted for by the layer-cake renormalization and the LPA with the
initial local potential taken from the cluster calculation (e.g., by
setting the momenta at the vertices to zero).  Within this approach the
low-${\bf k}$ region will shrink with the growing cluster radius $L_c$
which would diminish the LPA contribution thus making the calculation
more accurate because the larger cluster part should converge to the exact
solution. Besides, smaller $k_c$ means smaller renormalized interactions
which will make LPA more reliable which is exact to the first order in the
interactions \cite{1984}.  Thus, with the use of this hybrid cluster/LPA
approach the results obtained can be validated without resort to the
high-temperature expansions or MC simulations. The feasibility of the
approach is supported by the fact that in the purely cluster approach
good convergence could be seen in several cases with the use of small
easily manageable clusters \cite{tokar_new_1997,tan_topologically_2011}.
Finally, it is pertinent to note that besides making the LPA-based
approach self-contained, the hybrid technique would enable dealing with
the short-range cluster interactions that appear in the {\em ab initio}
theory of alloys \cite{ducastelle,Zunger2004}. This opens a possibility
of developing a theory which would make possible a realistic description
of the first- and the second-order phase transitions in lattice systems.

Another deficiency of the LPA is that it does not reproduce correctly the
critical exponents. This problem is not very severe because the exponents
are meaningful only asymptotically close to the critical point where the
renormalized Hamiltonian simplifies, acquires a universal form and, most
importantly, becomes of a weak-coupling kind in 3D case \cite{wilson}. So
the layer-cake renormalization can be stopped at sufficiently large
value of $t_c$ and the remaining renormalization performed with rigorous
perturbative techniques \cite{free_params2RG,free_params2RG2}. The
necessary Feynman diagrams with $\tilde{G}$ (\ref{G-g}) can be reduced
to the known expressions by separating $\tilde{G}$ into $G$ and $g$.
The advantage of this approach would be that all necessary input
parameters \cite{free_params2RG,free_params2RG2} will be known to a good
accuracy from the SC-LPA solution at $t_c$.

It should be noted that even in its simplest form the SC-LPA can be
useful in interpreting experimental data on the critical behavior. As was
shown in section \ref{beta-brass}, the non-universal contributions can
considerably distort the observed value of the exponent $\beta$ and SC-LPA
can describe the deviation of $\beta_{eff}$ from $\beta$ reasonably well,
as comparison with the exact MC simulations for the SC case shows (see
figure \ref{fig4}). Qualitatively similar deviations were observed
in experiments on the ordering in beta-brass \cite{beta-brass2016} but
they were larger than in our calculations. While theoretically the value
of $\beta$ for the Ising universality class is known with the accuracy
from three to four significant digits \cite{RG2002review}, in some fits
in \cite{beta-brass2016} even the first digit in $\beta_{eff}$ did not
agree with the RG value. Quantitative explanation of this discrepancy
requires knowledge of the ordering potential for beta-brass which can
be obtained either in {\em ab initio} calculations or from the diffuse
scattering data \cite{ducastelle,Zunger2004,us-ordering-potentials}.
Such explanation, besides resolving the issue with the value of $\beta$
in beta-brass, would also test the ability of the SC-LPA to quantitatively
describe non-universal behavior at finite distances from critical points
in realistic situations, thus extending the use of the RG method beyond
the immediate vicinity of the critical point.
\section*{Acknowledgment}
I would like to express my gratitude to Hugues Dreyss\'e for support
and encouragement.
\appendix\setcounter{section}{0}
\section{\label{ini}Initial condition in the spin models}
The spin models differ from the general $n$-vector case in that the
length of vectors ${\bf s}_i$ is fixed, for example, is equal to unity
\cite{n-vector-models}.  In the functional integral representation
(\ref{Z}) this can be taken into accounted by the product of the Dirac
delta-functions as
\begin{equation}
	e^{-H_I}=\prod_i\delta({\bf s}_i^2-1)
	=\lim_{u_4\to\infty}\left({\frac{u_4}{\pi}}\right)^{N/2}
	e^{-u_4\sum_i({\bf s}_i^2-1)^2}
	\label{delta}
\end{equation}
where the second equality shows that the spin Hamiltonian (\ref{H})
formally corresponds to $\phi^4$ model with infinitely strong
interaction. Equation (\ref{LPA}) would be difficult to integrate
numerically in this case so partial exact renormalization described in
section \ref{partial} is in order.

The $n=1$ case is trivial
\begin{equation}
	u^{(1)}({\bf x},t_0)
	=\frac{x^2}{2t_0}-\ln\cosh \frac{x}{t_0} + \mbox{(f.i.t.)}
\label{n=1}
\end{equation}
where $x=|{\bf x}|$ and (f.i.t.) stands for $x$-independent terms. 

For $n>1$ in the $O(n)$ symmetric case the integral in
(\ref{f_ini}) is convenient to calculated in hyperspherical
coordinates. Choosing the direction of ${\bf x}$ along the first axis
${\bf x}=(x\cos\theta,0,0,\dots,0)$ one gets \cite{n-sphere}
\begin{equation}
	e^{-u^{(n)}({\bf x},t_0)}
{\;\propto\;}e^{-\frac{x^2}{2t_0}}\int_0^{\pi}
e^{\frac{x}{t_0}\cos\theta}\sin^{n-2}\theta\,d\theta.
\label{theta}
\end{equation}
The cases $n=2$ and $n=3$ are given by
\begin{eqnarray}
u^{(2)}({\bf x},t_0)
&=&\frac{x^2}{2t_0}-\ln I_0\left(\frac{x}{t_0}\right) 
+ \mbox{(f.i.t.)}\nonumber\\
u^{(3)}({\bf x},t_0)
&=&\frac{x^2}{2t_0}-\ln \left(\frac{t_0}{x}\sinh\frac{x}{t_0}\right) 
+ \mbox{(f.i.t.)}
	\label{n=2,3}
\end{eqnarray}
where $I_0$ is the modified Bessel function of the first kind.

Explicit expressions for $n>3$ is not considered in the main text so
we only note that at large $n>8$ a three-term recurrence relation for
integrals in (\ref{theta}) can be established so additionally only $n=4-8$
integrals will need to be calculated explicitly for $n>3$.
\section{\label{x-y}Transformed RG equation}
In our case the Legendre transform suggested in \cite{x-yLegendre}
(see also \cite{local_potential}) should be modified as
\begin{eqnarray}
	\label{y-x}
y_\sigma({\bf x},t)&=&x_\sigma-(t-c)u_{x_\sigma}({\bf x},t)\\
	\label{v-u}
	v({\bf y},t)&=&u({\bf x},t)-\frac{1}{2}(t-c)\sum_\sigma u_{x_\sigma}^2({\bf x},t)
\end{eqnarray}
where $\sigma=1,\dots,n$ and $c$ is an arbitrary constant. Here the
independent variables are ${\bf x}$ and $t$ and the local potential is
$u$. Our aim is to use these relations to re-write (\ref{LPA}) in terms
of ${\bf y}$ and $t$ for the transformed potential $v$. To this end we
first differentiate Eqs.\ (\ref{y-x}) and (\ref{v-u}) with respect to
$x_{\sigma^\prime}$:
\begin{eqnarray}
	\label{y-x_diff}
\frac{\partial y_\sigma }{\partial x_{\sigma^\prime}}&=&
\delta_{\sigma\sigma^\prime}-(t-c)u_{x_\sigma x_{\sigma^\prime}}\\
	\label{v-u_diff}
 v_{y_\sigma }\frac{\partial y_\sigma }{\partial x_{\sigma^\prime}}
 &=&u_{x_{\sigma^\prime}} -(t-c)u_{x_\sigma }u_{x_\sigma x_{\sigma^\prime}}.
\end{eqnarray}
(summation over repeated subscripts is assumed).  Substituting
(\ref{y-x_diff}) into (\ref{v-u_diff}) after some rearrangement one
arrives at a linear system
\begin{equation}
	[\delta_{\sigma {\sigma^\prime}}-(t-c)u_{x_\sigma x_{\sigma^\prime}}]
	(v_{y_{\sigma^\prime}}-u_{x_{\sigma^\prime}})=0.
	\label{Mv-u}
\end{equation}
Because the matrix in this equation is not singular in general case, it
follows that for all $\sigma=1,\cdots,n$
\begin{equation}
	v_{y_\sigma }=u_{x_\sigma }.
	\label{v=u_diff}
\end{equation}
Differentiation of this with respect to $x_{\sigma^\prime}$ and using
(\ref{y-x_diff}) gives
\begin{equation}
	[\delta_{\sigma\kappa}+(t-c)v_{y_{\sigma}y_\kappa}]
	u_{x_\kappa x_{\sigma^\prime}}
	=v_{y_{\sigma}y_{\sigma^\prime}}.
	\label{uxx-vyy}
\end{equation}
As is seen, by the matrix inversion all $u_{x_\sigma x_{\sigma^\prime}}$
in the LPA equation can be expressed in terms of $v_{y_\kappa
y_{\kappa^\prime}}$.

Finally, differentiating Eqs.\ (\ref{y-x}) and (\ref{v-u}) by $t$ and
using Eq.\ (\ref{v=u_diff}) one gets
\begin{equation}
	u_t+\frac{1}{2}\sum_{\sigma}u_{x_{\sigma}}^2=v_t
	\label{ut-vt}
\end{equation}
(note the difference with equation (15) in \cite{x-yLegendre} where
the second term on the l.h.s.\ is absent) so that (\ref{LPA}) can be
written as
\begin{equation}
	v_t=\frac{1}{2}p(t)u_{x_\sigma x_\sigma}
	=\frac{1}{2}p(t)\nabla_{\bf x}^2 u
	\label{LPAv}
\end{equation}
where the r.h.s.\ should be expressed in terms of $v_{y_\sigma
y_{\sigma^\prime}}$ with the use of (\ref{uxx-vyy}). 
\subsection{$O(n)$ symmetric case} 
In a fully $O(n)$ symmetric case (\ref{uxx-vyy}) can be solved 
explicitly as follows. Introducing notation
\begin{equation}
	v({\bf y},t)=w(q,t)\mbox{\ where\ }q={\bf y}^2/2
\label{q}
\end{equation}
one finds
\begin{equation}
v_{y_{\sigma^\prime} y_\sigma }=\delta_{\sigma^\prime\sigma}w_q
+y_{\sigma^\prime} y_\sigma w_{qq}.
\label{v_yiyj}
\end{equation}
Now denoting the matrix in (\ref{uxx-vyy}) as $\hat{M}$ with the use of 
(\ref{v_yiyj}) one gets
\begin{eqnarray}
&&M_{\sigma\kappa}=\delta_{\sigma\kappa}+(t-c)v_{y_\sigma y_{\kappa}}\nonumber\\
&&=[1+(t-c)w_q]\left[\delta_{\sigma\kappa}
+y_\sigma y_{\kappa}\frac{(t-c)w_{qq}}{1+(t-c)w_q}\right]
\label{M}
\end{eqnarray}
Substituting this in (\ref{uxx-vyy}) and solving for 
$u_{x_\sigma x_{\sigma^\prime}}$ one arrives at the expressions that eliminates 
$u$ and ${\bf x}$ on the r.h.s.\ of (\ref{LPAv}) as
\begin{equation}
	\nabla^2_{\bf x} u =\frac{(n-1)w_q}{1+(t-c)w_q}
+\frac{w_q+2qw_{qq}}{1+(t-c)(w_q +2qw_{qq})}.
	\label{u_xixi}
\end{equation}
\subsection{$n=1$ case}
In the case $n=1$ the introduction of $q$ is superfluous because
directly from (\ref{uxx-vyy}) one gets \cite{x-yLegendre} 
\begin{equation}
	u_{xx}=\frac{v_{yy}}{1+(t-c)v_{yy}}.
	\label{uvn=1}
\end{equation}
Substituting this into (\ref{LPAv}) one obtains equation (\ref{the_eq2})
of the main text.  From (\ref{y-x}) and (\ref{v-u}) it is seen that the
most convenient choice of the arbitrary constant is $c=t_0$ so that
there is no need in the Legendre transform of the initial variables.
Now rewriting (\ref{uvn=1}) as
\begin{equation}
	\frac{1}{u_{xx}}=\frac{1}{v_{yy}}+t-t_0
	\label{u_1-v_1}
\end{equation}
it can be seen that if at some point $t\not=t_0$ $u_{xx}\to\infty$ $v_{yy}$
remains finite: $v_{yy}=-1/(t-t_0)$.

Substituting (\ref{u_1-v_1}) into (\ref{chi}) one finds the expression
for the inverse susceptibility in the transformed variables
\begin{equation}
	\chi^{-1}=\frac{1+\Delta t v_{yy}^R}{1-t_0v_{yy}^R}r.
	\label{chi-1}
\end{equation}


\section*{References}
\begin{thebibliography}{10}
\expandafter\ifx\csname url\endcsname\relax
  \def\url#1{\texttt{#1}}\fi
\expandafter\ifx\csname urlprefix\endcsname\relax\def\urlprefix{URL }\fi
\expandafter\ifx\csname href\endcsname\relax
  \def\href#1#2{#2} \def\path#1{#1}\fi

\bibitem{ducastelle}
F.~Ducastelle, Order and Phase Stability in Alloys, North-Holland, Amsterdam,
  1991.

\bibitem{tokar_new_1997}
V.~I. Tokar, A new cluster method in lattice statistics, Comput. Mater. Sci. 8
  (1997) 8--15.

\bibitem{maier_quantum_2005}
T.~Maier, M.~Jarrell, T.~Pruschke, M.~H. Hettler, Quantum cluster theories,
  Rev. Mod. Phys. 77 (2005) 1027--1080.

\bibitem{tan_topologically_2011}
T.~L. Tan, D.~D. Johnson, Topologically correct phase boundaries and transition
  temperatures for {Ising} {Hamiltonians} via self-consistent coarse-grained
  cluster-lattice models, Phys. Rev. B 83 (2011) 144427.

\bibitem{wilson}
K.~G. Wilson, J.~Kogut, The renormalization group and the $\epsilon$ expansion,
  Phys. Rep. 12 (1974) 75--199.

\bibitem{berges_non-perturbative_2002}
J.~Berges, N.~Tetradis, C.~Wetterich, Non-perturbative renormalization flow in
  quantum field theory and statistical physics, Phys. Rep. 363~(4) (2002) 223
  -- 386.

\bibitem{le_guillou_zinn-justin}
J.~C. Le~Guillou, J.~Zinn-Justin, Critical exponents from field theory, Phys.
  Rev. B 21 (1980) 3976--3998.

\bibitem{RG2002review}
A.~Pelissetto, E.~Vicari, Critical phenomena and renormalization-group theory,
  Phys. Rep. 368 (2002) 549--727.

\bibitem{BAGNULS200191}
C.~Bagnuls, C.~Bervillier, Exact renormalization group equations: an
  introductory review, Physics Reports 348~(1) (2001) 91--157, renormalization
  group theory in the new millennium. II.
\newblock \href
  {http://dx.doi.org/https://doi.org/10.1016/S0370-1573(00)00137-X}
  {\path{doi:https://doi.org/10.1016/S0370-1573(00)00137-X}}.

\bibitem{latticeRG2010}
T.~Machado, N.~Dupuis, From local to critical fluctuations in lattice models: A
  nonperturbative renormalization-group approach, Phys. Rev. E 82 (2010)
  041128.
\newblock \href {http://dx.doi.org/10.1103/PhysRevE.82.041128}
  {\path{doi:10.1103/PhysRevE.82.041128}}.

\bibitem{beta-brass2016}
A.~Madsen, J.~Als-Nielsen, J.~Hallmann, T.~Roth, W.~Lu, Critical behavior of
  the order-disorder phase transition in $\ensuremath{\beta}$-brass
  investigated by x-ray scattering, Phys. Rev. B 94 (2016) 014111.
\newblock \href {http://dx.doi.org/10.1103/PhysRevB.94.014111}
  {\path{doi:10.1103/PhysRevB.94.014111}}.

\bibitem{elliott_theory_1974}
R.~J. Elliott, J.~A. Krumhansl, P.~L. Leath, The theory and properties of
  randomly disordered crystals and related physical systems, Rev. Mod. Phys. 46
  (1974) 465--543.

\bibitem{gamma_exp}
V.~I. Tokar, A new series expansion in lattice statistics, Phys. Lett. A 110
  (1985) 453--456.

\bibitem{masanskii_method_1988}
I.~V. Masanskii, V.~I. Tokar, Method of $\gamma$-expansions in the electronic
  theory of disordered alloys, Theor. Math. Phys. 76~(1) (1988) 747--757.
\newblock \href {http://dx.doi.org/10.1007/BF01029433}
  {\path{doi:10.1007/BF01029433}}.

\bibitem{local_potential}
C.~Bervillier, Revisiting the local potential approximation of the exact
  renormalization group equation, Nucl. Phys. B 876 (2013) 587.

\bibitem{1984}
V.~I. Tokar, A new renormalization scheme in the {Landau-Ginzburg-Wilson}
  model, Phys. Lett. A 104 (1984) 135--139.

\bibitem{arXiv19}
V.~I. Tokar, \rm Calculation of non-universal thermodynamic quantities within
  self-consistent non-perturbative functional renormalization group approach
  (2019) arXiv:1904.10338.

\bibitem{hori_approach_1962}
S.~Hori, An approach to a relativistic strong coupling theory, Nucl. Phys. 30
  (1962) 644--663.

\bibitem{vasiliev1998}
A.~N. Vasiliev, Functional Methods in Quantum Field Theory and Statistical
  Physics, Gordon and Breach, Amsterdam, 1998.

\bibitem{lieb2001analysis}
E.~Lieb, M.~Loss, Analysis, {CRM} Proceedings \& Lecture Notes, American
  Mathematical Society, Providence, RI, 2001.

\bibitem{nicoll_exact_1976}
J.~F. Nicoll, T.~S. Chang, H.~E. Stanley, Exact and approximate differential
  renormalization-group generators, Phys. Rev. A 13 (1976) 1251--1264.

\bibitem{golner}
G.~R. Golner, Nonperturbative renormalization-group calculations for continuum
  spin systems, Phys. Rev. B 33 (1986) 7863--7866.
\newblock \href {http://dx.doi.org/10.1103/PhysRevB.33.7863}
  {\path{doi:10.1103/PhysRevB.33.7863}}.

\bibitem{IVANCHENKO1990100}
Y.~Ivanchenko, A.~Lisyanskii, A.~Filippov, New small rg parameter, Physics
  Letters A 150~(2) (1990) 100 -- 104.
\newblock \href
  {http://dx.doi.org/https://doi.org/10.1016/0375-9601(90)90258-P}
  {\path{doi:https://doi.org/10.1016/0375-9601(90)90258-P}}.

\bibitem{maxwell_construction}
A.~Parola, D.~Pini, L.~Reatto, First-order phase transitions, the {M}axwell
  construction, and the momentum-space renormalization group, Phys. Rev. E 48
  (1993) 3321--3332.

\bibitem{caillol_non-perturbative_2012}
J.-M. Caillol, The non-perturbative renormalization group in the ordered phase,
  Nuclear Physics B 855 (2012) 854--884.

\bibitem{x-yLegendre}
T.~Morris, Equivalence of local potential approximations, J. High Energy Phys.
  0507 (2005) 027.

\bibitem{Jelitto1969609}
R.~J. Jelitto, The density of states of some simple excitations in solids, J.
  Phys. Chem. Solids 30~(3) (1969) 609--626.

\bibitem{special_points}
H.~J. Monkhorst, J.~D. Pack, Special points for brillouin-zone integrations,
  Phys. Rev. B 13 (1976) 5188--5192.

\bibitem{Zunger2004}
V.~Blum, A.~Zunger, Mixed-basis cluster expansion for thermodynamics of bcc
  alloys, Phys. Rev. B 70 (2004) 055108.

\bibitem{us-ordering-potentials}
I.~V. Masanskii, V.~I. Tokar, T.~A. Grishchenko, Pair interactions in alloys
  evaluated from diffuse-scattering data, Phys. Rev. B 44 (1991) 4647--4649.
\newblock \href {http://dx.doi.org/10.1103/PhysRevB.44.4647}
  {\path{doi:10.1103/PhysRevB.44.4647}}.

\bibitem{lsode}
K.~Radhakrishnan, A.~C. Hindmarsh, Description and use of {LSODE}, the
  {L}ivermore solver for ordinary differential equations, Tech. Rep.
  UCRL-ID-113855, LLNL (December 1993).

\bibitem{bcc-fcc-diamond-Kc}
P.~Lundow, K.~Markström, A.~Rosengren, The ising model for the bcc, fcc and
  diamond lattices: A comparison, Phil. Mag. 89~(22--24) (2009) 2009--2042.
\newblock \href {http://dx.doi.org/10.1080/14786430802680512}
  {\path{doi:10.1080/14786430802680512}}.

\bibitem{n-vector-models}
P.~Butera, M.~Comi, Extension to order $\ensuremath{\beta}{}^{23}$ of the
  high-temperature expansions for the spin-$\frac{1}{2}$ ising model on simple
  cubic and body-centered cubic lattices, Phys. Rev. B 62 (2000) 14837--14843.

\bibitem{talapov_M(t)}
A.~L. Talapov, H.~W.~J. Bl{\"o}te, The magnetization of the 3d ising model, J.
  Phys. A 29 (1996) 5727.

\bibitem{liu_fisher89}
A.~J. Liu, M.~E. Fisher, The three-dimensional ising model revisited
  numerically, Physica 156A (1989) 35--76.

\bibitem{free_params2RG}
C.~Bagnuls, C.~Bervillier, Nonasymptotic critical behavior from field theory at
  $d=3$: The disordered-phase case, Phys. Rev. B 32 (1985) 7209--7231.

\bibitem{free_params2RG2}
C.~Bagnuls, C.~Bervillier, D.~I. Meiron, B.~G. Nickel, Nonasymptotic critical
  behavior from field theory at \textit{d=3}. {II. T}he ordered-phase case,
  Phys. Rev. B 35 (1987) 3585--3607.

\bibitem{n-sphere}
\em {S}pherical volume element, \url{https://en.wikipedia.org/wiki/N-sphere}.

\end{thebibliography}
\end{document}